# Off-centered Pb interstitials in PbTe

**Sungjin Park, Jongho Park, Byungki Ryu,* and SuDong Park**

Energy Conversion Research Center, Korea Electrotechnology Research Institute

12, Bulmosan-ro 10beon-gil, Seongsan-gu, Changwon-si, 51543, Republic of Korea

**Abstract**

In this work, we calculate the defect properties of low-symmetry Pb interstitials in PbTe using first-principles density-functional theory calculations. We break the symmetry imposed on on-centered interstitial defects and show that the lowest ground state of Pb interstitial defects is off-centered along the [111] directions. Due to the four multi-stable structures with low defect formation energies, the defect density of Pb interstitials is expected to be ~5.6 times larger than previous predictions when PbTe is synthesized at 900 K. In contrast to the on-centered Pb-interstitial, the off-centered Pb interstitials in PbTe can exhibit long-range lattice relaxation toward [111] direction beyond distance of 1 nm, indicating the potential formation of weak local dipoles. This result provides an alternative explanation for the *emphanitic anharmonicity* of PbTe.

* byungkiryu@keri.re.kr.

## I. INTRODUCTION

Thermoelectric effects enable direct energy conversion between thermal and electrical energies.[1] Thermoelectric conversion efficiency can be estimated with the dimensionless figure of merit $ZT = \alpha^2\sigma T/(\kappa_{elec}+\kappa_{latt})$, where $\alpha$ is the Seebeck coefficient and $\sigma$ is the electrical conductivity, $\kappa_{elec}$ and $\kappa_{latt}$ are the electronic and lattice thermal conductivity, respectively, and T is the absolute temperature.[2] Since a large ZT can lead to higher thermoelectric efficiency, reducing the lattice thermal conductivity and optimizing the thermoelectric power factor ($\alpha^2\sigma$) has been key strategies for developing materials with high thermoelectric performance.[1,3,4]

PbTe-based materials have high thermoelectric performance for middle temperatures applications up to the hot side temperature of 800 K.[5–11] With Na or Bi doping, suitable charge carriers are doped and the Fermi level is positioned near the optimal band edge positions, resulting in the optimization of the power factor.[12–15] Alloying or doping with extrinsic Ag, Sb, CdTe, MgTe, MnTe, SrTe, EuTe, or $Ag_2Te$ phases leads to low lattice thermal conductivity near ~1 W/m/K.[5,6,8,9,16–18] As a result, many high ZT PbTe-based materials have been developed.[5,10,17–20]

While further enhancement of ZT of PbTe has been investigated in alloying or doping studies, its intrinsic nature has also been considered to be the origin of high thermoelectric performance. Its complex low-energy band structure with non-parabolicity and high valley degeneracy is known to be responsible for the band convergence and high thermoelectric power factor $\alpha^2\sigma$.[7] Meanwhile, the anharmonic phonon structure has been considered as the origin of low thermal conductivity. While there are no off-centered Pb atoms in PbTe lattice,[21–23] there are large thermal displacements for Pb atoms at high temperatures, indicating a large phonon anharmonicity.[22,24–26] More recently, PbTe has been reported an *emphanitic anharmonicity* *behavior*, which refers to the configurational entropy driven by the formation of local dipoles. The space and time averaged structures are at the high-symmetry on-centered lattice position.[27] However, studies using single crystal x-ray diffuse scattering analysis with ab initio molecular dynamics have also revealed that the formation of local dipoles extends over a few unit cells. In addition, the time evolution is estimated to be ~10 picoseconds. This *emphanitic behavior* of local dipole fluctuations at high temperatures is different from the normal phase transition observed in GeTe, $Cu_2Te$, or $Ag_2Te$, where the low-symmetry globally polarized phases are stabilized at below the phase transformation temperature.[28–32]

Meanwhile, there is a lack of understanding of the intrinsic defect nature in PbTe. Recent first-principles calculations have revealed that PbTe has several intrinsic defects as the charged defect formation energies of intrinsic defects are small (~0 to 2 eV).[33–36] The point intrinsic defect picture describes well the electrical properties of binary phases, where under the Pb-rich condition, the material is known to have an n-type conduction behavior due to the shallow Te vacancy ($V_{Te}$) donor, while the Te-rich condition leads to a p-type conduction by the hole generation due to the shallow Pb vacancy ($V_{Pb}$) acceptor. However, there is still a lack of understanding of the relation between the existence of intrinsic defects and the PbTe lattice. Our recent observation of lattice volume expansion of PbTe under Pb-excess condition cannot be explained by previous results for the formation of point intrinsic defects.[36] Although interstitial defects can enlarge the lattice volume of PbTe, the most stable defect is not the Pb interstitial ($Pb_{Int}$) but the $V_{Te}$. To address the discrepancy between the theoretically predicted defect formation energy and the experimentally observed enlarged lattice parameter, the defect cluster model has been used to explain the lattice expansion.[36] Although the non-equilibrium synthesis of material can lead to the formation of a defect cluster, the defect formation energy difference between $Pb_{Int}$ and $V_{Te}$ is still high. Thus, there is still a lack of understanding of the interaction between the intrinsic defects and the PbTe lattice. Moreover, previous investigations have been based on the well-defined high-symmetry configuration, which neglects the possible interaction between intrinsic defects and the host PbTe lattice.

To further understand the formation of intrinsic defects in Pb-rich PbTe, we revisit the charged defect formation energies of intrinsic defects in binary PbTe, especially for $Pb_{Int}$. As the PbTe lattice is reported to be soft,[22] we carefully investigate possible low-symmetry configurations for vacancy, interstitial, and antisite defects. The positively charged $V_{Te}^{2+}$ defect state is the most stable intrinsic defect in Pb-rich PbTe, which is responsible for the n-type conduction. Meanwhile, $Pb_{Int}$ defects are the next most stable defects and the formation energy difference between the high-symmetry $V_{Te}^{2+}$ and $Pb_{Int}^{2+}$ defect configurations is about 0.55 eV. By breaking the symmetry imposed on high-symmetry configurations, the difference can be lowered by ~26 meV when the interstitial is displaced in the direction of one of the four nearest neighbor (NN) Te or in other equivalent directions. Moreover, the previously reported on-centered $Pb_{Int}$ site is found to be unstable in our results. Vacancies and antisite defects are also able to be off-centered. However, throughout this study, we focus on $Pb_{Int}$ defects because they can enlarge the lattice volume unlike the other type of defects such as $V_{Te}$. Beyond off-centered interstitial, we also find the Pb–Pb dimer interstitial configuration with higher defect formation energy than the on-centered one. On the other hand, in contrast to the off-centered structure, the on-centered interstitial and Pb–Pb dimer type defects are unstable or have saddle point configurations. However, due to the symmetry of their supercells, the structure can be frozen theoretically at these defect states. The off-centered $Pb_{Int}$ significantly affects the atomic positions in the host lattice. As the interaction is long-range, the atomic relaxation is found at a distance of ~1 nm from the defect site. Moreover, most of Pb and Te in the host PbTe are displaced along the $[\bar{1}\bar{1}\bar{1}]$ and $[111]$ directions to form a rhombohedral-like structure.

Our results indicate that, while perfect PbTe has a cubic rock-salt structure, the existence of low-symmetry interstitial defects can alter the local lattice structure due to the ferroelectric-like instability with weak local dipole formation, similar to that of GeTe. Furthermore, as each off-centered $\mathrm{Pb_{Int}}$ can have four different variant configurations, we expect that the $\mathrm{Pb_{Int}}$ in PbTe can form a local ferroelectric-like domain with different variant directions. Due to computational costs, we do not investigate the domain size. However, we believe that if such variant regions are generated, they might be responsible for intrinsically low thermal conductivity of binary PbTe by forming the phonon blocking barrier at the boundaries between $\mathrm{Pb_{Int}}$-induced ferroelectric-like domains, in addition to the strong phonon anharmonic nature of perfect PbTe.

## II. COMPUTATIONAL METHOD

We performed first-principle density functional theory (DFT) calculations[37,38] using the Vienna Ab initio simulation package (VASP).[39,40] We used a plane-wave basis set with an energy cutoff of 320 eV, the Perdew-Burke-Ernzerhof (PBE) exchange correlational functional,[41,42] and the projector augmented wave (PAW) pseudopotentials.[43,44] For the k-space integration in the Brillouin zone, we used a $3 \times 3 \times 3$ $\Gamma$-centered k-point mesh grid. To consider the relativistic effect of heavy elements, we employed spin-orbit interaction (SOI).[44]

We also used the calculated lattice parameter of 6.5758 Å for PbTe, which was obtained by the Murnaghan fitting.[45,46] Note that the calculated lattice parameter is slightly larger by 1.7% than the experimentally reported value of 6.462 Å.[47] This small lattice overestimation is the well-known error of generalized gradient approximation (GGA) calculations. As PbTe has one of the largest lattice parameters, the lattice overestimation can affect the defect structural relaxation. It was previously found experimentally that the lattice parameter can change up to ~2% with temperature changes from 0 to 900 K.[22] Thus, our optimized lattice parameter can also be applied for the high-temperature behavior of PbTe although it is larger than the 0 K lattice parameter of PbTe.

To investigate the defect structure of PbTe, a 128-atom FCC supercell was used. We considered all possible intrinsic point defects, i.e., interstitials, vacancies, and antisites. To make a defective supercell, one additional atom was added to the supercell for the interstitials, one original atom inside the supercell was subtracted from the supercell for the vacancies, and one original cation (or anion) inside the supercell was replaced with a counterpart anion (or cation) for the antisites. Then, we relaxed the defect structures until the remaining forces were smaller than $10^{-3}$ eV/Å. Note that the low atomic force criterion is critical to determine the atomic structures of defective PbTe due to long-range interactions between atoms in PbTe.[48] In particular, we explored various low-symmetry configurations of intrinsic defects in the supercell, by checking the configuration stability between perturbed structures. Finally, we categorized the interstitial defect configurations into stable, unstable, or saddle.

We investigated properties of point intrinsic defects with various charge states, i.e., 2+, neutral, and 2−. The formation energy ($\mathrm{E_{Form}}[\mathrm{D}^q]$) was calculated using

$$\mathrm{E_{Form}}[\mathrm{D}^q] = \mathrm{E_{Tot}}[\mathrm{D^q}] - \mathrm{E_{Tot}}[\mathrm{PbTe}] + \sum_i \Delta \mathrm{n}_i \mu_i + q(\mathrm{E_{VBM}} + \mathrm{dV_{loc}}), \text{[33,49]} \quad (1)$$

where $\mathrm{E_{Tot}}[\mathrm{D}^{q}]$ is the total energy of the defective PbTe supercell of the charge state of $q$ and $\mathrm{E_{Tot}}[\mathrm{PbTe}]$ is the total energy of pristine PbTe. In Eq. (1), for a specific atom $i$ inside the supercell, $\Delta \mathrm{n}_i$ and $\mu_i$ represent the number change of a specific atom (Pb or Te) inside the supercell and its atomic chemical potential, respectively. $\mathrm{E_{VBM}}$ is the valence band maximum (VBM) of the host PbTe and $\mathrm{dV_{loc}}$ is a local potential correction arising in the finite charged supercell calculations.[33]

## III. RESULTS AND DISCUSSION

Firstly, we revisit the charged defect formation energies of high-symmetry defect configurations in PbTe. Figs. 1(a) and 1(b) show the charged defect formation energies in PbTe under Pb- and Te-rich conditions, respectively. $\mathrm{V_{Te}}$, $\mathrm{Pb_{Int}}$, and Pb antiste at the Te site ($\mathrm{Pb_{Te}}$) are major defects in Pb-rich PbTe. Under the Pb-rich condition, the formation energy is the lowest for $\mathrm{V_{Te}}$, followed by $\mathrm{Pb_{Int}}$ and $\mathrm{Pb_{Te}}$. The Te antisite at the Pb site ($\mathrm{Te_{Pb}}$) and Te interstitial ($\mathrm{Te_{Int}}$) defects have high formation energies, which are larger than 1.75 eV. On the other hand, under the Te-rich condition, the lowest defect formation energy is found for $\mathrm{Te_{Pb}}$ and $\mathrm{V_{Pb}}$. These results are generally consistent with the previous calculations.[34–36] Note that the PBE band gap is calculated to be 0.099 eV, which is smaller than the experimentally observed band gap of 0.3–0.4 eV due to the band gap underestimation problem in DFT calculations. Thus, the band gap correction can change the stability order between defects, especially for Te-rich cases. However, for the Pb-rich case, the defect stability is less sensitive to the Fermi level position. Thus, for Pb-rich off-stoichiometric PbTe, the defect physics will be less sensitive to the calculation setting or choice of exchange-correlation functionals.

Under the Pb-rich condition, previously predicted defect formation energies[34,35] were inconsistent with the observed lattice expansion in PbTe.[36,50] It can be seen in Fig. 1(a) that $V_{Te}$ is the most dominant intrinsic defect under the Pb-rich condition. However, experimental results.[36,50] have reported that $Pb_{Int}$ can be the most possible intrinsic defect under Pb-rich conditions because the lattice parameter of off-stoichiometric PbTe increases with increasing Pb excess concentration ($N_{Pb}^{excess}$). In particular, Lee et al. reported that the lattice parameter of PbTe increases to ~0.3% when $N_{Pb}^{excess}$ increases to ~8%.[36] However, the existence of $V_{Te}$ inside PbTe hardly explains the lattice expansion of Pb-excess PbTe because the vacancy defects, in general, reduce the lattice volume. Considering that the experimental results are accurate, we assume that $Pb_{Int}$ defects may have been responsible for the increase of the lattice parameter and they can be considered as dominant intrinsic defects inside Pb-rich PbTe. Thus, all possible configurations of $Pb_{Int}$ inside the PbTe lattice are investigated with symmetry breaking.

Fig. 2(a) shows the rock-salt structure of pristine PbTe with a lattice parameter of 6.5758 Å, with the black and gold balls corresponding to Pb and Te inside the PbTe. In the primitive PbTe unit cell, there are two basis positions, i.e., Pb at (0 0 0) and Te at (1/2 1/2 1/2). Pb is surrounded by six NN Te atoms and Te has six NN Pb atoms. In addition, the bond length of Pb–Te is calculated to be 3.2879 Å. Here we call a small $Pb_4Te_4$ cubic in the PbTe as a *subcubic*, denoted by the blue dashed line in Fig. 2(a). Furthermore, the PbTe conventional cubic unit cell contains eight units of the *subcubic*. Although each *subcubic* is not a unit cell, they are geometrically equivalent due to the point symmetry of the PbTe lattice. Therefore, when searching for single-defect configurations of low-symmetry $Pb_{Int}$ defects, we only consider defects in a single *subcubic*.

To find potential local minima and metastable states, we look for all possible irreducible positions of $Pb_{Int}$. There are three distinct high-symmetry positions for interstitial in the supercell, i.e., the bond center of the Pb–Te bond, the plane center of the *subcubic*, and the body center of the *subcubic* (BC). To break the symmetry of the defective *subcubic*, we consider the following configurations perturbed from the high-symmetry positions. For $Pb_{Int}$ at the bond center, $Pb_{Int}$ can be perturbed in the NN Pb direction, NN Te direction, and two normal directions of the bond such as the direction to BC and the direction to the adjacent plane center. Thus, $Pb_{Int}$ at the bond center has only four perturbed configurations. Besides, in the case of $Pb_{Int}$ at the plane center, $Pb_{Int}$ can be displaced in the direction of two types of NN atoms such as Pb and Te. In addition, it can be disturbed in the normal direction of the plane equivalent to the direction to BC from the plane center and in the direction perpendicular to the BC direction equivalent to the adjacent bond center direction from the plane center. Therefore, $Pb_{Int}$ at the plane center also has only four perturbed configurations. Furthermore, for $Pb_{Int}$ at BC, $Pb_{Int}$ can be dislocated in four directions, i.e., the directions to two distinguishable NN atoms such as Pb and Te, the direction to the NN plane center of the *subcubic*, and the direction to the NN bond center of the *subcubic*. Note that, due to the geometry of the *subcubic*, four NN Pb (or Te) directions from the BC are equivalent, six NN plane-center directions from the BC have the same symmetry, and twelve NN bond-center directions from the BC are an equivalent geometrical direction. Consequently, even $Pb_{Int}$ at BC has only four perturbed configurations. As a result, considering the above $Pb_{Int}$ positions, which include the high-symmetry positions and symmetry-broken low-symmetry positions, we perform the structural relaxation for 15 irreducible $Pb_{Int}$ positions.

We find three critical configurations from the five BC-related $Pb_{Int}$ configurations. For $Pb_{Int}$ positions related to the bond center and the plane center of the *subcubic*, the defect formation energies of the 2+ charge state are larger by 1.6 eV and 0.7 eV than the $Pb_{Int}$ located at the BC-related high-symmetry position ($Pb_{Int}^{on}$), respectively. In addition, based on their stability and structural features, we select three local extrema such as $Pb_{Int}^{on}$, BC-related $Pb_{Int}$ with a slight movement in the direction of one of NN Te atoms ($Pb_{Int}^{off}$), and BC-related $Pb_{Int}$ with a far movement toward one NN Pb to form a Pb–Pb dimer ($Pb_{Int}^{dim}$). Note that $Pb_{Int}^{on}$ is not a stable defect, but $Pb_{Int}^{off}$ is the ground state configuration for $Pb_{Int}$. Furthermore, $Pb_{Int}^{dim}$ is the saddle configuration and has a larger formation energy than $Pb_{Int}^{on}$.

Figs. 2(b)–2(d) show the schematic atomic structures of three important interstitial defects, i.e., $Pb_{Int}^{on}$, $Pb_{Int}^{off}$, and $Pb_{Int}^{dim}$, respectively, where the red balls indicate the schematic atomic positions of $Pb_{Int}$ for each configuration. As shown in Fig. 2(b), $Pb_{Int}^{on}$ is located at the center of the *subcubic*, which is one of the high-symmetry positions. This high-symmetry position has four NN Pb and four NN Te atoms. In Figs. 2(c) and 2(d), the symmetry-breaking induced structures are shown, where one is slightly displaced in the direction of one NN Te along $[111]$, but the other is far displaced toward Pb along $[\bar{1}\bar{1}\bar{1}]$. Note that the latter is stable only when $Pb_{Int}$ forms a symmetric Pb–Pb dimer structure, which is a saddle point. Otherwise, the $Pb_{Int}$ atom is relaxed back toward the $Pb_{Int}^{off}$ position.

In Fig. 3, we investigate the effects of symmetry breaking on the defect stability of $Pb_{Int}$ such as $Pb_{Int}^{on}$, $Pb_{Int}^{off}$, and $Pb_{Int}^{dim}$. All three defects are shallow donors as they are stable in the 2+ charge state for all the Fermi level range. $Pb_{Int}^{off}$ is the most stable defect because it has the lowest formation energy. Although the formation energy difference is small, there is a significant structural difference, which we discuss below. The difference of the formation energy between $Pb_{Int}^{off}$ and $Pb_{Int}^{on}$ is 26 meV when the defect charge state is 2+. Note that in the defect charge state of 2+, the formation energy difference is smaller than the difference between their total energies due to the effect of the local potential correction in Eq. (1). For the neutral charge state, the formation energy difference is reduced to 4 meV. When they are negatively charged, no energy difference is found between off-centered and on-centered ones. This charge state-dependent energetics imply that $\mathrm{Pb}_{\mathrm{Int}}^{\mathrm{off}}$ only appears when the defects are positively charged. On the other hand, $\mathrm{Pb}_{\mathrm{Int}}^{\mathrm{dim}}$ has much larger formation energy than the on-centered $\mathrm{Pb}_{\mathrm{Int}}^{\mathrm{on}}$. In the 2+ charge state, the energy difference is 498 meV. The energy difference is slightly reduced to 382 meV and 241 meV in the charge state of neutral and 2−, respectively. We emphasize that our finding is the first report of off-centered interstitial defects in PbTe, compared to the previously reported high-symmetry configuration.[34,35] Since there are four possible defect positions with lower defect formation energies, there can be more interstitial defects compared to the previous prediction. If we assume that the defect is generated at the temperature of 900 K and the formation energy difference is 26 meV, then the ratio of the defect density of off-centered ones to that of on-centered ones is 5.6. Thus, symmetry breaking can enhance the defect density by ~460 %.

Next, we investigate the effects of $\mathrm{Pb}_{\mathrm{Int}}$ on the atomic structure of PbTe. To measure the structural change of each atom in the supercell, we define the structural relaxation parameter by defect D of atom *i* as

$$\mathrm{R}_i(\mathrm{D}) := \left|\vec{r}_i - \vec{r}_i^{(0)}\right|, \tag{2}$$

where $\vec{r}_i$ is the position vector of the atom *i* in the supercell after structural relaxation by the $\mathrm{Pb}_{\mathrm{Int}}$ defect, and $\vec{r}_i^{(0)}$ is the position vector of atom *i* before structural relaxation, i.e., the original atomic position in the pristine PbTe. We also define the distance parameter $d$ of atom

$i$ from the defect D as

$$d_{\mathrm{D}-i} := |\vec{r}_i - \vec{r}_d|. \tag{3}$$

For the distance change, we also compute the distance $d_{D-i}^{(0)}$ from the ideal defect position to the ideal atomic position before relaxation.

Table I shows the short-range structural relaxation of neighboring host atoms near the $\mathrm{Pb_{Int}}$ defects, where the distance $d_{D-i}^{(0)}$ is smaller than or equal to 7.165 Å. Before structural relaxation, there are 4 Pb NNs and 4 Te NNs, 12 Pb next NNs (NNNs) and 12 Te NNNs, and 12 Pb next NNNs (NNNNs) and 12 Te NNNNs at distances of 2.847 Å, 5.452 Å, and 7.165 Å for $\mathrm{Pb_{Int}^{on}}$, respectively. After structural relaxation, the distances from the defect to the neighbors change. Note that the distances to Pb NNs and Te NNs increase regardless of the kind of $\mathrm{Pb_{Int}}$. In the case of $\mathrm{Pb_{Int}^{on}}$, after structural relaxation, the distance to Pb NNs is slightly larger than that to Te NNs, i.e., 3.349 Å and 3.108 Å, respectively. The structural change difference may be understood by the electrostatic interactions between charges of defects and local environments. The structural relaxation behavior of $\mathrm{Pb_{Int}^{off}}$ is distinguished from that of on-centered defects since the distances from $\mathrm{Pb_{Int}^{off}}$ to NNs are separated into two groups. After symmetry breaking, the $\mathrm{Pb_{Int}}$ is off-centered. The distance from the defect to one of four NN Pb (or Te) atoms increase while the distance to the other three atoms decrease. Thus, there are three slightly shorter bonds and one slightly longer bond. Although the structural relaxation parameters $\mathrm{R}_i(\mathrm{D})$ of NNNs and NNNNs are smaller than that of NN atoms, the structural change does not vanish even for the host atoms beyond the NN shells. For example, the distances from the defect to 12 NNN atoms or 12 NNNN atoms are not equal in the supercell involving the off-centered defect. The symmetry-breaking phenomenon is also clearly observed for the NNN and NNNN atomic shells of both Pb and Te atoms. In addition, the values of $d_{\mathrm{D}-i}$ for NNNs and NNNNs are significantly larger than those of NNs. Thus, although structural relaxation is the largest for the atoms in the NN shells, the structural change does not vanish even for the atoms far from the off-centered defect.

Next, we investigate the long-range effect of $\mathrm{Pb_{Int}}$ in PbTe. Fig. 4 shows the relation between the structural relaxation parameter and the defect distance. We find that a long-range interaction between $\mathrm{Pb_{Int}}$ and PbTe is stronger when the defect symmetry is lowered; the interaction range by $\mathrm{Pb_{Int}^{off}}$ exceeds 1 nm. For $\mathrm{Pb_{Int}^{on}}$, the $\mathrm{R}_i(\mathrm{D})$ is not negligible for $d_{\mathrm{D}-i} =$ 8.626 Å, while it begins to vanish at $d_{\mathrm{D}-i} =$ 9.728 Å. However, for $\mathrm{Pb_{Int}^{off}}$, there is a significant long-range interaction in our defect supercell. $\mathrm{R}_i(\mathrm{D})$ is still 0.079 Å for $d_{\mathrm{D}-i} =$ 13.346 Å. From Table I and Fig. 4, we can conclude that the off-centered defect structure is different from the on-centered one. The supercell containing $\mathrm{Pb_{Int}^{dim}}$ also shows clear differences in structural relaxation than the other configurations as shown in Figs. 2 and 4. Due to $\mathrm{Pb_{Int}^{dim}}$ at the Pb site, the structural relaxation of the supercell is significantly larger for the shorter distance than the others. However, the relaxation rapidly decreases compared to the off-centered configuration. In addition, the average $\mathrm{R}_i(\mathrm{D})$ of the supercell involving $\mathrm{Pb_{Int}^{dim}}$ with respect to Pb and Te is smaller than that of $\mathrm{Pb_{Int}^{off}}$ but is slightly larger than that of $\mathrm{Pb_{Int}^{on}}$. Note that, although the atomic structure involving $\mathrm{Pb_{Int}^{dim}}$ is also a low-symmetry structure, it has a higher symmetry than the off-centered one.

We analyze the effect of charge state on the structural change of $\mathrm{Pb_{Int}^{off}}$ by computing $\Delta d_{D-i}$ defined as

$$\Delta d_{D-i} := \frac{d_{D-i}^{\mathrm{off}} - d_{D-i}^{\mathrm{on}}}{d_{D-i}^{\mathrm{on}}} \times 100. \tag{4}$$

The results in Fig. 5 show that the structural relaxation of Pb is much larger than that of Te regardless of the charge state except for the 2− charge state. This is consistent with the results in Fig. 4, where the introduction of $\mathrm{Pb_{Int}}$ has greater effects on Pb than Te in its supercell. In addition, it can be seen that as the charge state changes from positive to negative, the overall $\Delta d_{D-i}$ clearly decreases, indicating that the off-centered defect configuration is recovered to the on-centered configuration. It also indicates that $\mathrm{Pb_{Int}}$ has a greater effect on the atomic structure of the PbTe lattice as the charge state changes from negative to positive, which is consistent with the results of the charged defect formation energy, i.e., the energy lowered by the symmetry breaking is significant for the 2+ charge state and vanishes for the 2− state.

We then investigate the effect of $\mathrm{Pb_{Int}}$ in PbTe on lattice distortion by computing the atomic distances between each host Pb and its NNN Te since Pb or Te inside PbTe experiences structural changes by the $\mathrm{Pb_{Int}}$. Note that the NNN Te is in the $[111]$ direction or in other equivalent directions from each host Pb in PbTe. Because each supercell containing the three kinds of $\mathrm{Pb_{Int}}$ has different structural features, we expect that the strain effect of the corresponding $\mathrm{Pb_{Int}}$ is distinct. To verify this, we calculate the structural distortion parameter along the $[111]$ direction, defined as

$$\overline{d^{[111]}_{(\mathrm{Pb-Te})_{\mathrm{NNN}}}} := \frac{\sum d^{[111]}_{(\mathrm{Pb-Te})_{\mathrm{NNN}}}}{N^{\mathrm{Te}}_{\mathrm{NNN}}}, \tag{5}$$

for all atoms in each defective supercell, where $d^{[111]}_{(\mathrm{Pb-Te})_{\mathrm{NNN}}}$ is the distance from one Pb atom to its NNN Te atom in the host PbTe and $N^{\mathrm{Te}}_{\mathrm{NNN}}$ is the total number of NNN Te of all host Pb in the PbTe supercell. Note that pristine PbTe has a fixed $d^{[111]}_{(\mathrm{Pb-Te})_{\mathrm{NNN}}}$ of 5.695 Å. In Table II, all $\overline{d^{[111]}_{(\mathrm{Pb-Te})_{\mathrm{NNN}}}}$ of the supercell involving $\mathrm{Pb_{Int}}$ is smaller than that of pristine PbTe.

Furthermore, $\overline{d^{[111]}_{(\mathrm{Pb-Te})_{\mathrm{NNN}}}}$ for atoms far from the defects, i.e., $d_{D-i} >$ 4.886 Å, significantly decreases in the order of $\mathrm{Pb}^{\mathrm{on},2+}_{\mathrm{Int}}$, $\mathrm{Pb}^{\mathrm{dim},2+}_{\mathrm{Int}}$, and $\mathrm{Pb}^{\mathrm{off},2+}_{\mathrm{Int}}$. Note that, due to the stronger interaction between $\mathrm{Pb_{Int}}$ and the host atoms with $d_{D-i} \leq$ 4.886 Å, the case of including only NN atoms of $\mathrm{Pb_{Int}}$ achieves a smaller $\overline{d^{[111]}_{(\mathrm{Pb-Te})_{\mathrm{NNN}}}}$ than the case of excluding them. This is consistent with our finding in Table I that atoms in the NN shell of $\mathrm{Pb_{Int}}$ has larger position changes than the other atoms. Note that the distances between $\mathrm{Pb_{Int}}$ and its NN atoms for all supercells are smaller than 4.886 Å. Moreover, even for the case excluding NNs of $\mathrm{Pb_{Int}}$, $d^{[111]}_{(\mathrm{Pb-Te})_{\mathrm{NNN}}}$ decreases in the order of $\mathrm{Pb}^{\mathrm{on},2+}_{\mathrm{Int}}$, $\mathrm{Pb}^{\mathrm{dim},2+}_{\mathrm{Int}}$, and $\mathrm{Pb}^{\mathrm{off},2+}_{\mathrm{Int}}$, which reflects the size order of the average $\mathrm{R}_i(\mathrm{D})$ of Pb and Te inside the host PbTe (see Fig. 4). As a result, each Pb and its NNN Te in the host PbTe can form a weak local dipole, indicating that $\mathrm{Pb_{Int}}$-induced ferroelectric-like domain can occur inside the host supercell. In particular, $\mathrm{Pb}^{\mathrm{off},2+}_{\mathrm{Int}}$ may induce a stronger ferroelectric-like domain than the other interstitials.

From above results, we elucidate that the structural change by interstitial defects is significant for the atoms far from the defects. This structural change is significant for [111] direction which is the direction of off-centering of Pb interstitial defects. Finally, we investigate that the Pb interstitial defects induce ferroelectric-like phase transformation in our supercell of $Pb_{65}Te_{64}$. For this, we calculate the degree of the ferroelectric-like phase transformation in the PbTe with interstitial defects by computing the average structural change of Pb or Te atoms. Note that we do not allow the lattice dynamics of defective supercells, but only the atomic structural relaxations. Without interstitial defects, the Pb and Te atoms are at (0 0 0) and (1/2 1/2 1/2) positions, respectively. However, with interstitial defects, the average positions of the Pb and Te atoms are changed to (δ δ δ) and (1/2 1/2 1/2), respectively, where δ is 0.018 for the supercell of $Pb_{65}Te_{64}$, which is comparable to the $\delta = \sim 0.023$ of GeTe when a = 6.00 Å. We expect that, if we allow the lattice distortion, the structural distortion can be enhanced having larger δ for PbTe. Related to the *emphanisis* of PbTe,[22,23,27,51,52] the host Pb in pristine PbTe has a temporal or spatial average position at the rock-salt lattice point, but the existence of $\mathrm{Pb_{Int}}$ causes the host Pb in the host PbTe to be permanently out of its original position and at a position with a specific displacement along the direction of its NNN Te.

## IV. CONCLUSIONS

We have performed DFT calculations to find the low-symmetry off-centered Pb interstitial defects with lower defect formation energies. We find that the off-centered Pb interstitial is the multi-stable defect while the on-centered defect is unstable. We also find the saddle Pb–Pb dimer interstitial structure. Due to the lower defect formation energy of the multi-stable off-centered defect, we find that the defect density is larger than previous findings. From the structural analysis, we reveal that the structural distortion along the [111] direction is significant for the 2+ charge state, which is reduced in the neutrally and negatively charged states. In contrast to on-centered defects, off-centered defects show long-range structural relaxation effects, which might be responsible for the local rhombohedral phase transformation. We believe that intrinsic Pb interstitial off centering is another possible mechanism of *emphanitic anharmonicity* in PbTe.

## ACKNOWLEDGMENTS

This work was supported by KERI primary research program through the NST funded by the MSIT of the Republic of Korea (ROK): grant No. 20A01025. It was also supported by the KETEP and the MOTIE of the ROK: grant No. 20188550000290.

## Captions of Tables

Table I. Distributions of $d_{D-i}^{(0)}$, $\mathrm{R}_i(\mathrm{D})$, $d_{D-i}$, and the degeneracy (#) inside the supercell involving $\mathrm{Pb_{Int}}$ ($\mathrm{Pb_{Int}^{on}}$ and $\mathrm{Pb_{Int}^{off}}$) in the Pb shells (top of the table) and the Te shells (bottom of the table), where $d_{D-i}^{(0)} \leq 7.165$ Å.

Table II. $\overline{d_{(\mathrm{Pb-Te})_{\mathrm{NNN}}}^{[111]}}$ of host Pb atoms with $d_{D-i}$ less than 4.886 Å or greater than 4.886 Å in the supercells involving $\mathrm{Pb_{Int}}$ such as $\mathrm{Pb_{Int}^{on}}$, $\mathrm{Pb_{Int}^{off}}$, and $\mathrm{Pb_{Int}^{dim}}$.

Table I. Distributions of $d^{(0)}_{D-i}$, $\mathrm{R}_i(\mathrm{D})$, $d_{D-i}$, and the degeneracy (#) inside the supercell involving $\mathrm{Pb_{Int}}$ ($\mathrm{Pb^{on}_{Int}}$ and $\mathrm{Pb^{off}_{Int}}$) in the Pb shells (top of the table) and the Te shells (bottom of the table), where $d^{(0)}_{D-i} \leq 7.165$ Å.

| **Pb shell** | $\boldsymbol{d^{(0)}_{D-i}}$ | $\mathbf{Pb^{on}_{Int}}$ | | | $\mathbf{Pb^{off}_{Int}}$ | | |
|---|---|---|---|---|---|---|---|
| | | $\mathbf{R}_i(\mathbf{D})$ | $\boldsymbol{d_{D-i}}$ | # | $\mathbf{R}_i(\mathbf{D})$ | $\boldsymbol{d_{D-i}}$ | # |
| NN | 2.847 | 0.502 | 3.349 | 4 | 0.453 | 3.338 | 3 |
| | | | | | 0.511 | 3.346 | 1 |
| NNN | 5.452 | 0.049 | 5.5 | 12 | 0.043 | 5.438 | 3 |
| | | | | | 0.067 | 5.517 | 3 |
| | | | | | 0.096 | 5.532 | 6 |
| NNNN | 7.165 | 0.061 | 7.225 | 12 | 0.054 | 7.172 | 3 |
| | | | | | 0.096 | 7.216 | 6 |
| | | | | | 0.101 | 7.299 | 3 |
| **Te shell** | $\boldsymbol{d^{(0)}_{D-i}}$ | $\mathbf{Pb^{on}_{Int}}$ | | | $\mathbf{Pb^{off}_{Int}}$ | | |
| | | $\mathbf{R}_i(\mathbf{D})$ | $\boldsymbol{d_{D-i}}$ | # | $\mathbf{R}_i(\mathbf{D})$ | $\boldsymbol{d_{D-i}}$ | # |
| NN | 2.847 | 0.261 | 3.108 | 4 | 0.257 | 3.107 | 3 |
| | | | | | 0.298 | 3.108 | 1 |
| NNN | 5.452 | 0.061 | 5.512 | 12 | 0.042 | 5.487 | 3 |
| | | | | | 0.096 | 5.514 | 6 |
| | | | | | 0.089 | 5.538 | 3 |
| NNNN | 7.165 | 0.016 | 7.181 | 12 | 0.041 | 7.171 | 3 |
| | | | | | 0.057 | 7.183 | 6 |
| | | | | | 0.056 | 7.185 | 3 |

Table II. $\overline{d^{[111]}_{(\mathrm{Pb-Te})_{\mathrm{NNN}}}}$ of host Pb atoms with $d_{D-i}$ less than 4.886 Å or greater than 4.886 Å in the supercells involving $\mathrm{Pb_{Int}}$ such as $\mathrm{Pb_{Int}^{on}}$, $\mathrm{Pb_{Int}^{off}}$, and $\mathrm{Pb_{Int}^{dim}}$.

| **Supercell Type** | $\overline{d^{[111]}_{(\mathrm{Pb-Te})_{\mathrm{NNN}}}}$ **(Å)** | |
|---|---|---|
| | $d_{D-i} \leq$ **4.886** | $d_{D-i} >$ **4.886** |
| $\mathrm{Pb_{Int}^{on,2+}}$ | 5.519 | 5.670 |
| $\mathrm{Pb_{Int}^{off,2+}}$ | 5.521 | 5.630 |
| $\mathrm{Pb_{Int}^{dim,2+}}$ | 5.551 | 5.660 |

## Captions of Figures

FIG. 1. Formation energy ($E_{Form}$) of a high-symmetry on-centered single intrinsic defect in binary PbTe under (a) Pb-rich and (b) Te-rich conditions. The red and blue lines correspond to the data for Pb- and Te-related defects, where the solid, dotted, and dashed lines represent the result for interstitials ($Pb_{Int}$ and $Te_{Int}$), vacancies ($V_{Pb}$ and $V_{Te}$), and antisites ($Pb_{Te}$ and $Te_{Pb}$), respectively. The blue and yellow regions represent the valence and conduction band of PbTe, respectively. The white region between the two regions represents the energy gap of our PBE calculation ($E_{Gap}^{PBE}$) in PbTe. $E_{Fermi}$ and $E_{VBM}$ indicate the Fermi level and the VBM of our calculation, respectively.

FIG. 2. (a) Atomic structure of pristine PbTe in rock-salt structure, The blue dashed line cubic represents the *subcubic*. Schematic figures of $Pb_{Int}$ inside the PbTe lattice (b) at the high-symmetry on-centered position ($Pb_{Int}^{on}$), (c) at the off-centered position slightly moved in the direction of one NN Te ($Pb_{Int}^{off}$), and (d) at the position far moved toward the NN Pb to form a Pb−Pb dimer ($Pb_{Int}^{dim}$). The black and gold balls represent Pb and Te atoms inside PbTe, respectively. The plane displayed in panels (b), (c), and (d) is the purple plane of (a). In (b), (c), and (d), the red balls represent the atomic positions of $Pb_{Int}$ for each configuration. Note that the solid and dotted lines are just guides to the eye.

FIG. 3. Defect formation energy ($E_{Form}$) of $Pb_{Int}^{off}$, $Pb_{Int}^{on}$, and $Pb_{Int}^{dim}$ in PbTe. The x-axis represents the difference between the Fermi level ($E_{Fermi}$) and the valence band maximum energy ($E_{VBM}$) of PbTe. The y-axis represents the defect formation energy of $Pb_{Int}$ in PbTe. The red solid, blue dashed, and black dotted lines represent the $E_{Form}$ of $Pb_{Int}^{off}$, $Pb_{Int}^{on}$, and $\mathrm{Pb}_{\mathrm{Int}}^{\mathrm{dim}}$, respectively. The blue and yellow regions represent the valence and conduction band of PbTe, respectively. The white region between the blue and yellow regions represents the PBE energy gap.

FIG. 4. Effect of introducing $\mathrm{Pb}_{\mathrm{Int}}$ on $\mathrm{R}_i(\mathrm{D})$ of host atoms excluding its NN Pb and its NN Te inside PbTe in the charge state of 2+ for (a) $\mathrm{Pb}_{\mathrm{Int}}^{\mathrm{on}}$, (b) $\mathrm{Pb}_{\mathrm{Int}}^{\mathrm{off}}$, and (c) $\mathrm{Pb}_{\mathrm{Int}}^{\mathrm{dim}}$. The reference coordinate of $\mathrm{Pb}_{\mathrm{Int}}^{\mathrm{dim}}$ is (0, 0, 0), which is the center of the Pb−Pb dimer. The x-axis and y-axis represent $d_{D-i}$ and $\mathrm{R}_i(\mathrm{D})$ in each supercell, respectively. The red × and blue + represent the data for Pb and Te atoms, respectively. The red and blue dashed lines represent the average of $\mathrm{R}_i(\mathrm{D})$ for Pb and Te in each supercell, respectively.

FIG. 5. $\Delta d_{D-i}$ of each atom (Pb and Te) with respect to the charge state of the supercell: (a) 2+, (b) neutral, and (c) 2−. The x-axis and y-axis represent $d_{D-i}$ and $\Delta d_{D-i}$ in the supercell involving $\mathrm{Pb}_{\mathrm{Int}}^{\mathrm{off}}$, respectively. The red × and blue + represent the data of host Pb and host Te, respectively.

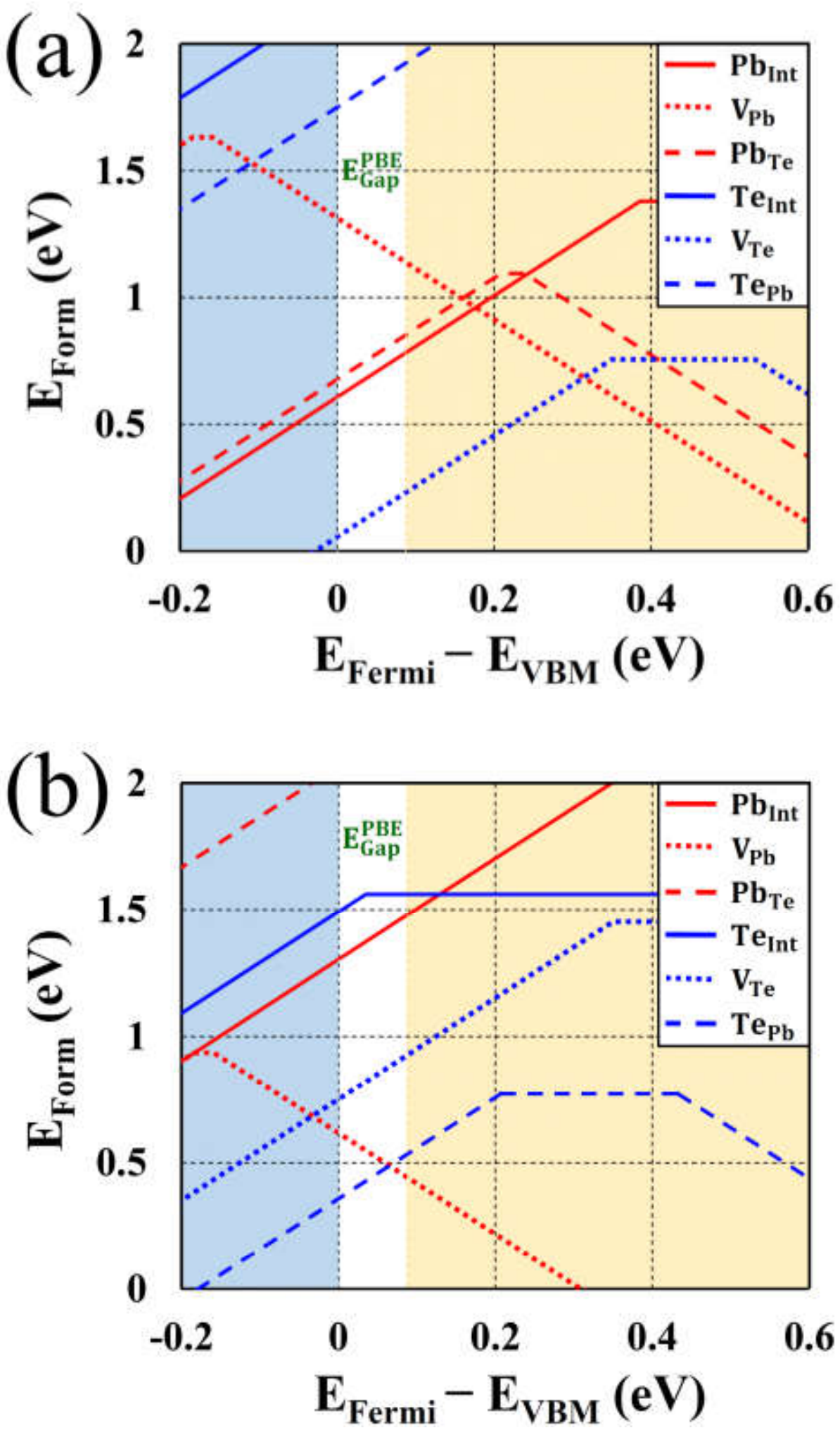

FIG. 1. Formation energy ($E_{Form}$) of a high-symmetry on-centered single intrinsic defect in binary PbTe under (a) Pb-rich and (b) Te-rich conditions. The red and blue lines correspond to the data for Pb- and Te-related defects, where the solid, dotted, and dashed lines represent the result for interstitials ($Pb_{Int}$ and $Te_{Int}$), vacancies ($V_{Pb}$ and $V_{Te}$), and antisites ($Pb_{Te}$ and $Te_{Pb}$), respectively. The blue and yellow regions represent the valence and conduction band of PbTe, respectively. The white region between the two regions represents the energy gap of our PBE calculation ($E_{Gap}^{PBE}$) in PbTe. $E_{Fermi}$ and $E_{VBM}$ indicate the Fermi level and the VBM of our calculation, respectively.

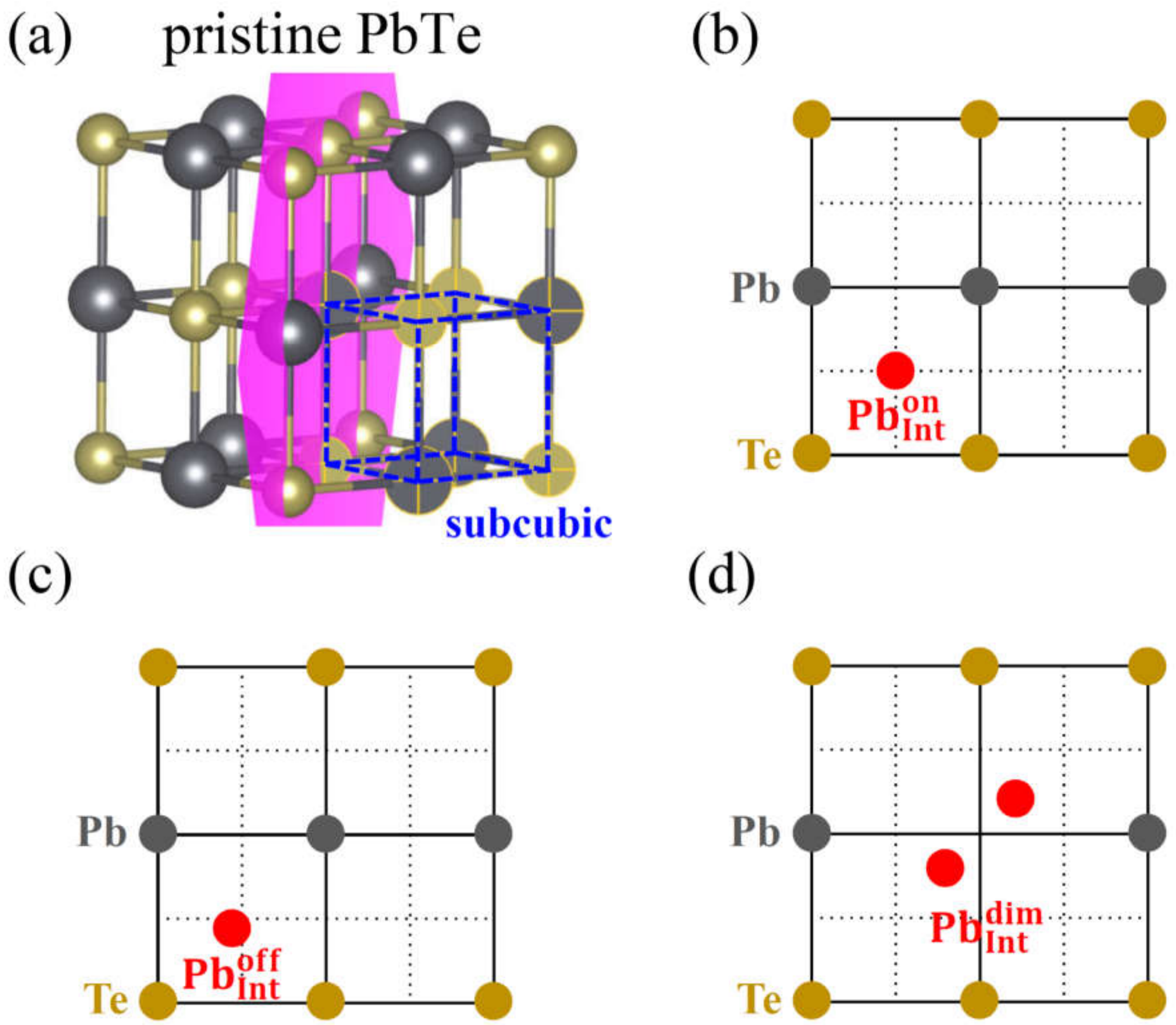

FIG. 2. (a) Atomic structure of pristine PbTe in rock-salt structure, The blue dashed line cubic represents the *subcubic*. Schematic figures of $Pb_{Int}$ inside the PbTe lattice (b) at the high-symmetry on-centered position ($Pb_{Int}^{on}$), (c) at the off-centered position slightly moved in the direction of one NN Te ($Pb_{Int}^{off}$), and (d) at the position far moved toward the NN Pb to form a Pb−Pb dimer ($Pb_{Int}^{dim}$). The black and gold balls represent Pb and Te atoms inside PbTe, respectively. The plane displayed in panels (b), (c), and (d) is the purple plane of (a). In (b), (c), and (d), the red balls represent the atomic positions of $Pb_{Int}$ for each configuration. Note that the solid and dotted lines are just guides to the eye.

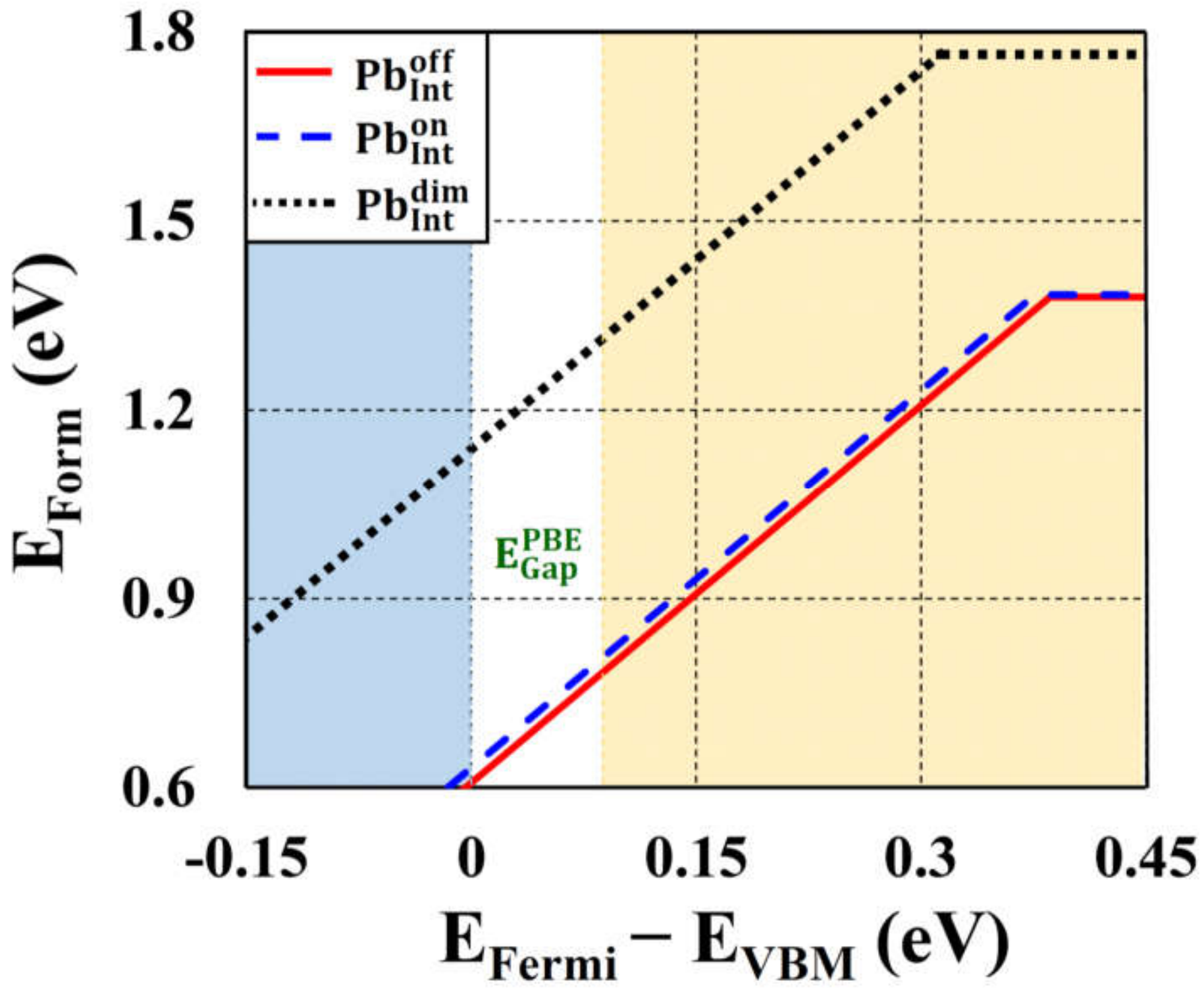

FIG. 3. Defect formation energy ($E_{Form}$) of $Pb_{Int}^{off}$, $Pb_{Int}^{on}$, and $Pb_{Int}^{dim}$ in PbTe. The x-axis represents the difference between the Fermi level ($E_{Fermi}$) and the valence band maximum energy ($E_{VBM}$) of PbTe. The y-axis represents the defect formation energy of $Pb_{Int}$ in PbTe. The red solid, blue dashed, and black dotted lines represent the $E_{Form}$ of $Pb_{Int}^{off}$, $Pb_{Int}^{on}$, and $Pb_{Int}^{dim}$, respectively. The blue and yellow regions represent the valence and conduction band of PbTe, respectively. The white region between the blue and yellow regions represents the PBE energy gap.

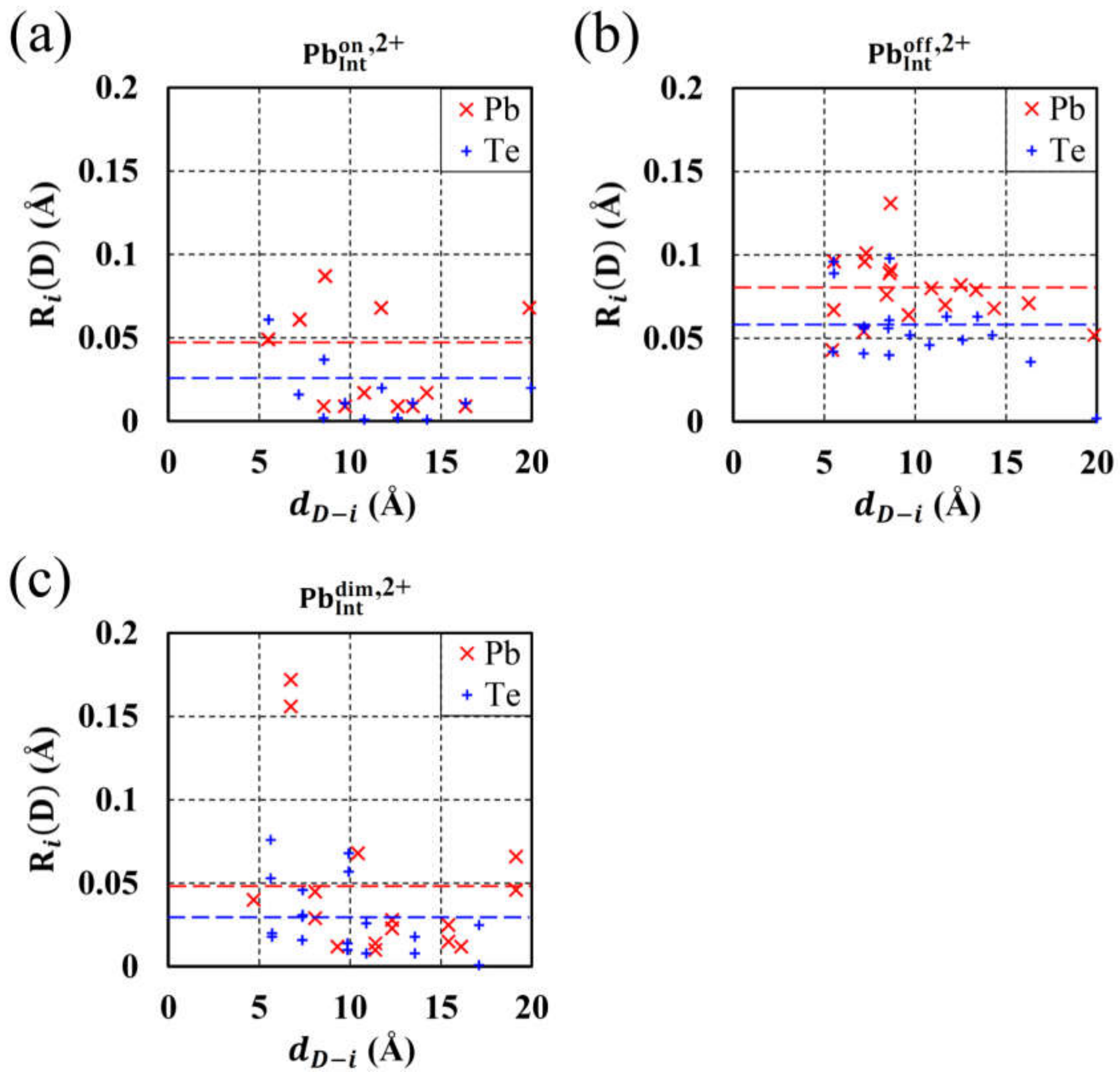

FIG. 4. Effect of introducing $Pb_{Int}$ on $R_i(D)$ of host atoms excluding its NN Pb and its NN Te inside PbTe in the charge state of 2+ for (a) $Pb_{Int}^{on}$, (b) $Pb_{Int}^{off}$, and (c) $Pb_{Int}^{dim}$. The reference coordinate of $Pb_{Int}^{dim}$ is (0, 0, 0), which is the center of the Pb−Pb dimer. The x-axis and y-axis represent $d_{D-i}$ and $R_i(D)$ in each supercell, respectively. The red × and blue + represent the data for Pb and Te atoms, respectively. The red and blue dashed lines represent the average of $R_i(D)$ for Pb and Te in each supercell, respectively.

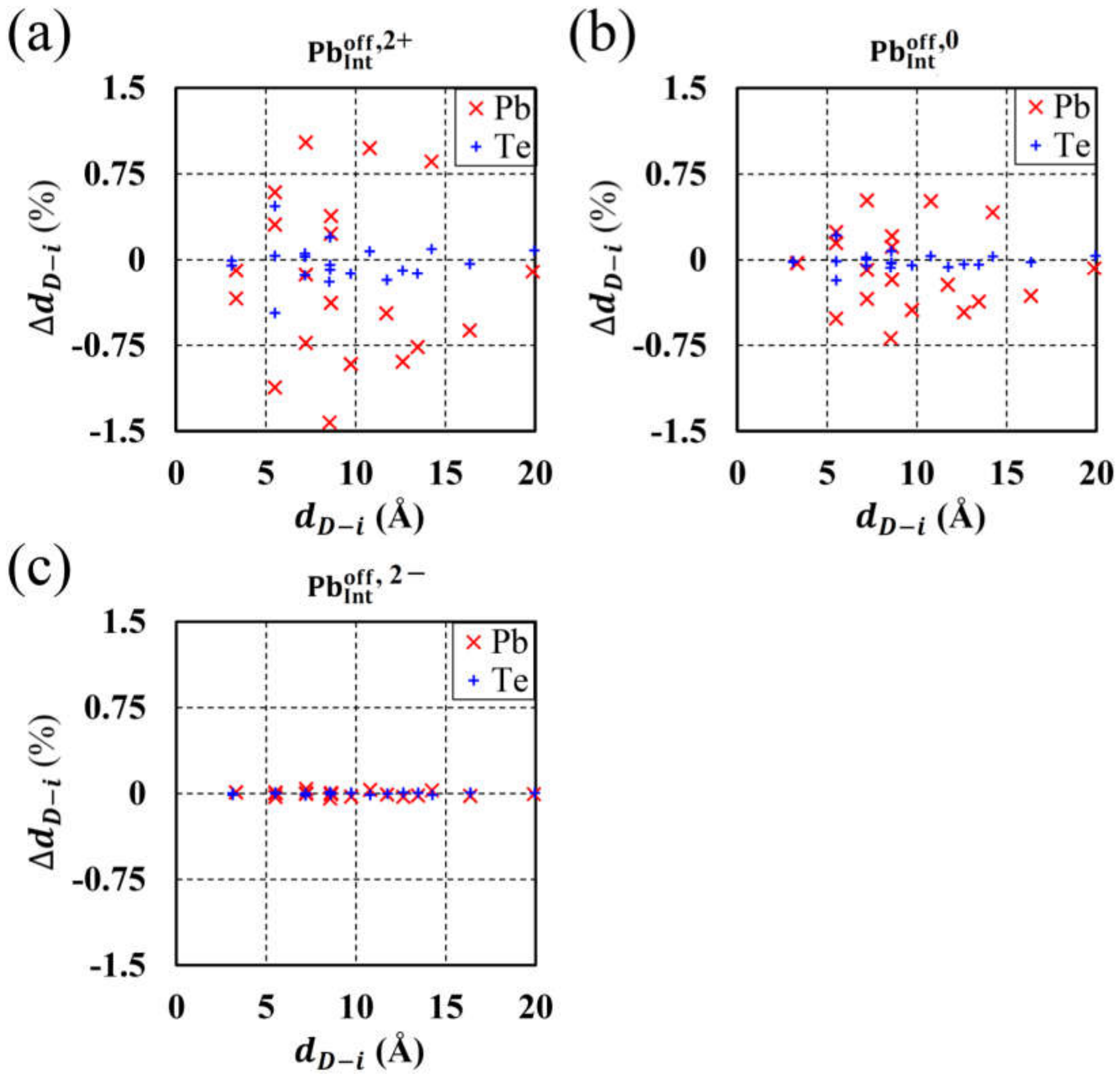

FIG. 5. $\Delta d_{D-i}$ of each atom (Pb and Te) with respect to the charge state of the supercell: (a) 2+, (b) neutral, and (c) 2−. The x-axis and y-axis represent $d_{D-i}$ and $\Delta d_{D-i}$ in the supercell involving $Pb_{Int}^{off}$, respectively. The red × and blue + represent the data of host Pb and host Te, respectively.